\documentclass[usegraphicx]{mn2e}
\begin{document}

\title[MOND and the lensing FP]{MOND and the Lensing Fundamental Plane:
No need for dark matter on galaxy scales}

\author[R.H. Sanders \& D.D. Land] {R.H.~Sanders and
 D.D. Land\\Kapteyn Astronomical Institute,
P.O.~Box 800,  9700 AV Groningen, The Netherlands}

 \date{received: ; accepted: }

\maketitle

\begin{abstract}
Bolton et al. (2007) have derived a mass-based fundamental
plane using photometric and spectroscopic observations of 36 
strong gravitational lenses.  The lensing allows a direct
determination of the mass-surface density and so avoids the
usual dependence on mass-to-light ratio.  We consider this
same sample in the context of modified Newtonian dynamics
(MOND) and demonstrate that the observed mass-based
fundamental plane coincides with the 
MOND fundamental plane determined previously for a set of
polytropic spheres chosen to match the observed range
of effective radii and velocity dispersions in elliptical
galaxies.  Moreover, the observed
projected mass within one-half an effective radius
is consistent with the mass
in visible stars plus a small additional component 
of ``phantom dark matter'' resulting from the MOND contribution
to photon deflection.
\end{abstract}

\section{Introduction}

Modified Newtonian dynamics (MOND) posits a fundamental
acceleration scale ($a_0$) below which the law of gravity
or inertia deviates from the Newtonian form (Milgrom 1983).
Therefore, a generic prediction is
that in systems where the internal acceleration, expressed
as a surface density, exceeds a critical
limit ($a_0/G$), there should be little discrepancy between the 
observable mass and the Newtonian dynamical mass.  In
traditional language, there should be no evidence for
dark matter within high surface brightness systems.  

Examples of
high surface brightness systems are globular clusters and
luminous elliptical galaxies.  It is well-known that
the standard Newtonian analysis of the internal kinematics 
of globular clusters  
yields mass-to-light ratios that are completely consistent
with normal stellar populations (e.g., Pryor et al. 1989).  
However, dark matter
proponents were surprised by the more recent analysis of planetary
nebulae kinematics in several luminous ellipticals which
implied a ``dearth of dark matter'' within the bright inner
regions of these galactic systems (Romanowsky et al. 2003).
This result is completely consistent with the
expectations of MOND (Milgrom \& Sanders 2003).

The galaxy scale gravitational lenses that produce multiple
images of background sources-- strong lenses-- are primarily in early
type galaxies: ellipticals or bulge dominated spirals \cite{koch04};
that is to say, high surface brightness systems. Because the
surface density of lenses necessary to produce multiple
images generally exceeds the MOND critical surface
density, a MOND corollary prediction is that the implied
mass surface density in strong gravitational lenses
should be generally consistent with that of
observable starlight, i.e., little or no dark matter in
strong, galaxy scale, gravitational lenses, at least
within the Einstein ring radius: ``what you see
is all there is.''

Recently, Bolton et al. (2007) have
considered the ``fundamental plane'' of early type galaxies
(the observed relation between effective radius, surface 
brightness and velocity dispersion) as defined by
a sample of 36 strong lens galaxies from the
Sloan Lens ACS.  The lensing analysis was combined with spectroscopic
and photometric observations of the individual lens galaxies
in order to generate a ``more fundamental plane'' based
upon mass surface density rather than surface brightness.
They found that this {\it mass-based} fundamental plane relation 
exhibits less scatter and is closer to the expectations
of the Newtonian virial relation than is the usual 
luminosity-based fundamental plane.  This presumably is due to the
absence of both an intrinsic scatter in M/L as well as its 
systematic variation with luminosity.  Below, we emphasise
that the implied
lensing M/L values within the Einstein ring radius do not require the
presence of a substantial component of dark matter, again consistent
with the expectations of MOND.

With MOND, the mass-velocity
dispersion ($M-\sigma$) relation (or its observational equivalent, 
the luminosity-based Faber-Jackson law), 
is, in some sense,
more fundamental than the fundamental plane.  In fact, near
isothermal, homologous objects exhibit a precise mass-velocity
dispersion relation (Milgrom 1984).
But, in order to match the observed
range of elliptical galaxy properties-- primarily the distribution
of the velocity dispersion vs. effective radius-- 
models for such systems must deviate from strict homology.
The necessary deviation yields a large scatter
in the predicted $M-\sigma$ law; none-the-less, the models 
define a narrow mass-based
fundamental plane when an additional parameter is included,
such as effective radius or surface density (Sanders 2000).  

Here, we compare this previously derived MOND fundamental
plane to the mass-based fundamental plane of Bolton et al.
and find that they are entirely consistent.  
Using the theoretical fundamental plane relation to
calculate the mass from the observed effective radius and velocity 
dispersion, it is found that the projected
MOND FP mass, presumably entirely baryonic, is proportional to the
observed lensing mass with a small offset
due to the effect of modified gravity on the deflection of photons
along the line of sight.  
Moreover, the implied MOND M/L-colour
relationship for these systems is entirely consistent with population
synthesis models.  In other words, we find that 
MOND requires no dark matter
on galaxy scale in lensing galaxies, in contrast to recent
claims (Ferreras et al. 2008).

\section{The More Fundamental Plane}

The fundamental plane of elliptical galaxies (Dressler et al. 1987,
Djorgovski \& Davis 1987)
is a scaling relationship involving the observed effective radius, $r_{eff}$,
the central line-of-sight velocity dispersion, $\sigma_0$, and, in 
the original
form, the mean surface brightness within an effective radius, $I$.
This is usually expressed as
$$\log(r_e)= a\log(\sigma)+ b\,\log(I) + d \eqno(1)$$
(it may alternatively be written as a relationship between total
luminosity, effective radius, and velocity dispersion).  The 
analysis of Sloan data on elliptical galaxies implies that
$a\approx 1.5$ and $b\approx -0.8$ (Benardi et al. 2003), 
while the expectations from the Newtonian
virial theorem, assuming homology and constant M/L, would
be $a=2$ and $b=-1$. The observed deviation from these expectations
is generally thought to be due to systematic deviations from
homology or from constant M/L.

The significant achievement of Bolton et al. was to use 
gravitational lensing in order to eliminate the dependence on
M/L.  The strong gravitational lensing provides the mass surface
density within the Einstein ring radius ($r_{ein}$); therefore surface
density, $\Sigma$, may be substituted for surface brightness in the 
above relationship. The Einstein ring radius in their sample is generally
about half the effective radius, so the surface density is taken
by Bolton et al. 
to be that within $r_{eff}/2$ (the correction from the $r_{ein}$ to
$r_{eff}/2$ is done either by assuming a mass distribution like that
of an isothermal sphere or that light traces mass).
This was then combined with spectroscopic and photometric observations
of the lens galaxies to generate a ``mass'' fundamental plane.
The resulting relation
exhibits considerably less scatter and the exponents are close
to the virial expectations: $a=1.86\pm 0.17$, 
$b=-0.93\pm 0.09$, $d=5.4\pm 0.9$ with the mass traces light
assumption (here the effective radius is in kpc, the velocity
dispersion in kms$^{-1}$ and the surface density in M$_\odot$ kpc$^{-2}$).

\begin{figure}
\resizebox{\hsize}{!}{\includegraphics{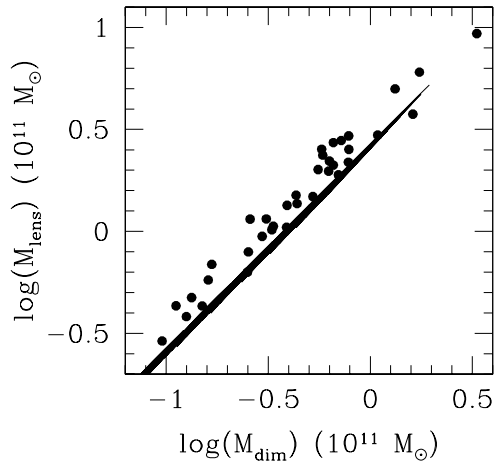}}
\caption[]{The points illustrate the mass-based fundamental
plane of Bolton et al. (2007) shown as a relation between
the observed projected mass within $r_{eff}/2$ and the dimensional
mass (eq.\ 2).  Also shown is the same for the MOND fundamental
plane as determined from 360 anisotropic high n polytropes
(Sanders 2000). Mass is given in units of $10^{11}$ M$_\odot$.} 
\label{}
\end{figure}

The Newtonian virial theorem implies that
the mass within radius $r$ is given by
$$M=c\sigma^2 r/G\eqno(2)$$
where $c$ is a constant determined by the structure of the object.
Therefore Bolton et al. also express their result by plotting the 
mass of the lens (within $r_{eff}/2$) against a ``dimensional''
mass variable given by $\sigma^2 r_{eff}/(2G)$ where $\sigma$ is
the mean line-of-sight velocity dispersion within $r_{eff}/2$.
The result of this is shown by the points in Fig. 1 which are
well fit by 
$$\log(M_{lens}) = \delta \log(M_{dim}) + C.\eqno(3)$$
with $\delta=0.98$ and $C\approx 0.6$.
Therefore, these observations are consistent with 
the Newtonian virial theorem and no variation of
the structure constant;
elliptical galaxies would appear to be quite homologous over
a wide range of mass.  Moreover, the lensing M/L values (r-band) 
for these galaxies range from about two to eight in solar units and
are somewhat higher than predicted by population synthesis models
(Fig. 3, upper panel).  This point is addressed further below, but,
overall, for early-type galaxies
M/L values in this range would not constitute compelling
evidence for dark matter within $r_{eff}/2$.

\section{The MOND fundamental plane}

By solving the structure equation,
Milgrom (1984) has demonstrated that, with MOND, isothermal
spheres with a fixed degree of anisotropy
have finite mass, an upper limit to the surface
density ($\approx a_0/G$) 
and exhibit a $M-\sigma^4$ relationship
(the basis of the Faber-Jackson law); 
therefore, such objects might constitute 
sensible models for elliptical galaxies.  However, the 
observed properties of ellipticals-- in particular, their
distribution and scatter
on the $\sigma$-$r_{eff}$ plane \cite{jor99,jfk95a,jfk95b}-- 
cannot be matched
by isothermal spheres or any strictly homologous class of objects.  
The MOND isothermal sphere is too inflated for
a given velocity dispersion-- the mean surface density within an
effective radius is too low.  In order
to match these observed properties of 
elliptical galaxies with MOND, it is necessary to
exploit other degrees of freedom:  models must deviate from
being strictly isothermal with a fixed degree of orbital
anisotropy.

In order to explore these possible degrees of freedom,
Sanders (2000) has considered polytropic spheres with
an anisotropy parameter, $\beta = 1-{\sigma_t}^2/{\sigma_r}^2$,
which varies systematically with radius
as in the Osipkov-Merritt models, $\beta = r^2/(r^2+{r_a}^2)$,  
($r_a$ is the anisotropy radius, Binney \& Tremaine 1987).
For a polytropic sphere of index $n$ the radial velocity dispersion-density
relationship is given by 
$${\sigma_r}^2 = A\rho^{1/n}\eqno(4)$$
where $A$ is a constant ($n\rightarrow \infty$ corresponds to the
isothermal sphere).  It was found that a range $12<n<16$ was sufficient
to match the the mean value and wide dispersion of effective
radius for a given velocity dispersion, provided that the
anisotropy radius also varies over the range $2<r_a/r_{eff}<25$.
The luminosity or mass density distribution within these objects 
is similar to that of
a Jaffe model (Jaffe 1982):  within roughly $r_{eff}$ the density
falls as $r^{-2}$ and beyond steepens to $r^{-4}$.
Spherically symmetric N-body calculations with MONDian
modified gravity (Bekenstein \& Milgrom 1984) demonstrate
that objects resembling such large $n$ polytropic spheres may actually
condense and recollapse out of the Hubble flow (Sanders 2008). 

Each model, characterised by a particular value of $n$
and $r_a/r_{eff}$, exhibits its own exact $M\propto \sigma^4$ relationship
but for all models combined there is a large dispersion in this
relation.  None-the-less, in spite of the dispersion in homology,
the models lie on a narrow fundamental plane
$$\log(r_e) = 2.0\,\log(\sigma_{e2}) - 1.06\,\log(\Sigma_{e2}) + 6.16
\eqno(5)$$
where $\sigma_{e2}$ and $\Sigma_{e2}$ are the velocity dispersion
and the average surface density respectively within $r_{eff}/2$
(the coefficients here differ slightly from those given by
Sanders because of scaling to $r_{eff}/2$).
This is seen to be quite close to the Newtonian virial expectation
and within the errors of the more fundamental plane defined
by the gravitational lenses.  In fact, both the MOND and
lensing fundamental
planes are well-approximated by the virial relation for 
Newtonian Jaffe models.

As above, the MOND fundamental plane relation may be
expressed in terms
of a MOND FP mass (projected within $r_{eff}/2$) as a function of
the dimensional mass unit $\sigma^2 r_{eff}/2G$.  This is shown by
the thick line in Fig.\ 1.  This is the original ensemble of 360 
anisotropic, polytropic models covering the range described above.
We see that it is almost coincident with the observations of
Bolton et al.

\begin{figure}
\resizebox{\hsize}{!}{\includegraphics{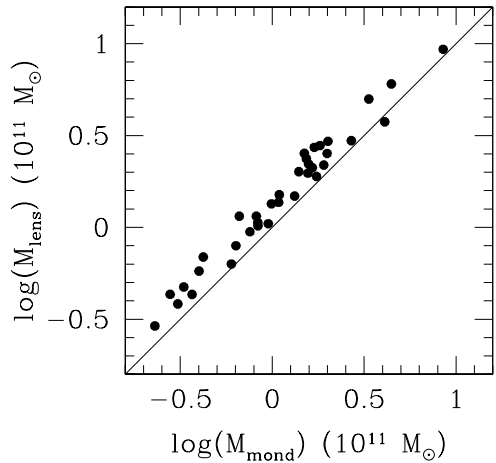}}
\caption[]{ The logarithm of the observed
lensing mass at $r_{eff}/2$ (Bolton et al. 2007) plotted against
the logarithm of the projected MOND FP mass within $r_{eff}/2$ calculated
via the MOND fundamental plane, eq.\ 5 (Sanders 2000).  The solid line
is the line of equality.  The 30\% offset is obvious.} 
\label{}
\end{figure}

We may also apply the MOND fundamental plane relation in order
to calculate the mass
of a lens galaxy, projected within $r_{eff}/2$, from the observed velocity 
dispersion and effective radius.  This calculated MOND FP mass is plotted
against the observed lensing mass in Fig.\ 2.  It is evident that
the MOND FP mass is closely proportional to the lens mass, but about
30\% less on average.  This is understandable because the MOND FP mass
is the total projected baryonic mass; the lensing mass, however, includes
a phantom dark matter contribution because MOND, or its relativistic
extension, TeVeS (Bekenstein 2004), 
provides extra deflection along the line-of-sight
\cite{mt01,zeal06}. 

\begin{figure}
\resizebox{\hsize}{!}{\includegraphics{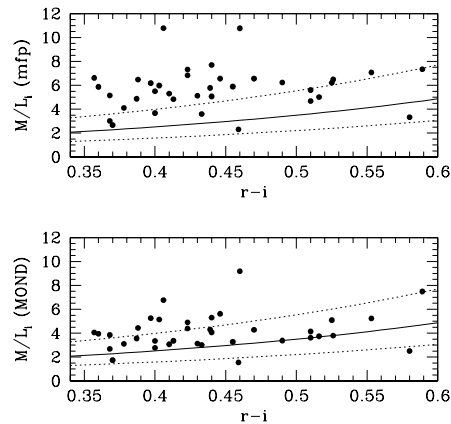}}
\caption[]{The upper panel shows the observed lensing M/L within
$r_{eff}/2$ in the $r$ band plotted against the $r-i$ colour.
The solid line is the prediction of the population synthesis models
of Bell et al. 2004 with the Kennecut or
Kroupa initial mass function.  The dashed lines show the 0.2 dex range
which is a realistic estimate in the uncertainty of the
estimated stellar M/L values given the uncertainty of
the initial mass function and the effects of metallicity.
The lower panel is the same for the MOND
M/L.} 
\label{}
\end{figure}

It is of interest to compare the MOND M/L values with
population synthesis models.  In Fig.\ 3 (lower panel)
the indicated M/L values
within $r_{eff}/2$ are plotted  against $r-i$ colour and compared
to the theoretical models of Bell et al. (2003).  The luminosities
have been determined from the observed Sloan magnitudes applying
both K- and evolutionary corrections (Poggianti 1997) and
reduced by a factor of about three to correspond
to that within $r_{eff}/2$.
We see that MOND M/L
values closely track the theoretical values.  The upper panel is
the same except that the M/L is that determined from the
projected lensing mass.  Again
the same trend is present although the offset discussed above
is evident.  

\section{The contribution of modified gravity to the deflection of
photons}

The current multi-field relativistic extensions of MOND, such as TeVeS,
are characterised by a physical metric which is distinct from the
Einstein, or gravitational, metric.  The transformation between
the two metrics is ``disformal'' and chosen such that the
relationship between the total weak field gravitational force and
the deflection of photons is identical to that of General
Relativity; specifically, the deflection angle is given by
$$\alpha = {2\over {c^2}}\int{g_\perp dz} \eqno(6)$$
where the integral is along the line of sight and 
$g_\perp$ is the perpendicular component of gravitational force.
The gravitational force includes not only the traditional
Newtonian force but also an additional `MOND' force (mediated by
a scalar field) which becomes dominant at low accelerations.

This MOND force may be viewed as arising from a phantom dark
halo.  The phantom halo begins to make a significant contribution
to the gravitational force and hence to the deflection of
photons near the MOND transition radius
$$r_t = \sqrt{GM/a_0}\eqno(7)$$
where $M$ is the true (baryonic) mass of the object.  In fact,
the phantom halo may exhibit an apparent shell of matter between 
1/2 $r_t$ up to $r_t$ depending upon the
form of the function which interpolates
between the MOND regime and the Newtonian regime \cite{ms08}.  
The relevance for the present discussion is
that we would expect to find some evidence of ``dark matter''
in strong gravitational lensing if the Einstein ring radius is
a significant fraction of the MOND transition radius;  the
larger that fraction, the larger the apparent contribution of
dark matter to the projected mass detected by strong gravitational
lensing.

The Einstein ring radius, for a point mass, is given by
$$r_{ein} = {{4GM}\over{c^2}}{{D_l D_{ls}}\over{D_s}} \eqno(8)$$
where $D_l$, $D_s$, and $D_{ls}$ are respectively the angular 
size distances from the observer to the lens, to the source and
from the lens to the source.  Combining eqs.\ 7 and 8 and
making use of the fact that $a_0=fcH_0$ where $f\approx 1/6$ 
we find that
$$r_{ein} = r_t F(z_l,z_s)\eqno(9)$$
where F is a function of the lens and source redshift and depends
upon the cosmological model. 

\begin{figure}
\resizebox{\hsize}{!}{\includegraphics{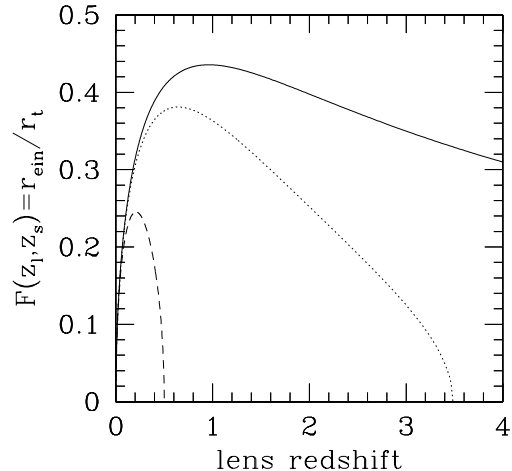}}
\caption[]{The Einstein radius in terms of the MOND transition
radius as a function of the lens redshift for three different
values of the source redshift:  $z_s =$ 0.5 (dashed curve), 3.5
 (dotted), 100 (solid).  This
is for the standard ``concordance'' cosmology.} 
\label{}
\end{figure}

This function $F(z_l,z_s)$ is shown in Fig.\ 4 as a function
of $z_l$ for three different values of the source redshift
($z_s = 0.5,\,3.5,\,100$).  This is all done for the standard
``concordance'' cosmology which the relativistic MOND cosmology
should  mimic to a close approximation.  For the lenses in the sample of
Bolton et al., $z_s\approx 0.5$ and $z_l \approx 0.2$ on average;
therefore, $F$ does not exceed about 0.25, and we might expect the
contribution of an apparent halo to lensing to be relatively small.
On the other hand for strong lenses considered by Treu
\& Koopmans (2004), $z_l\approx 1$ and $z_s\approx 3.5$ implying 
$F\approx 0.4$.  The expectation is that
there should be evidence for more ``dark matter''
in these high redshift systems, 
as, in fact, is reported by Treu \& Koopmans.

\begin{figure}
\resizebox{\hsize}{!}{\includegraphics{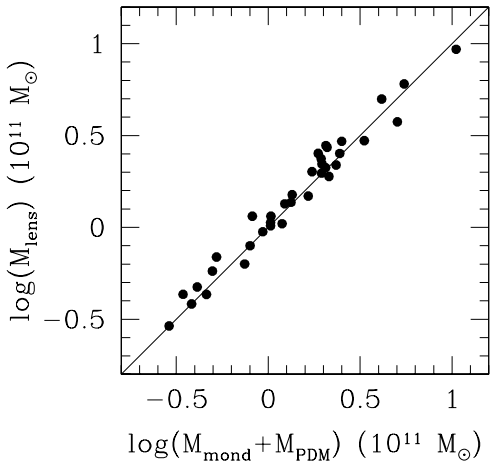}}
\caption[]{The observed lensing mass within $r_{eff}/2$ from
Bolton et al. plotted against the projected MOND FP mass including the
contribution of phantom dark matter ($M_{PDM}$.  This corrects Fig.\ 2
for the effect of modified gravity on gravitational lensing.} 
\label{}
\end{figure}

We have estimated the contribution of
modified gravity to lensing, in the form of projected phantom dark 
matter, for all of the objects in the sample of Bolton et al.
The objects have been modelled as isotropic Jaffe spheres with the
observed effective radius.  Given this density distribution, the
Jeans equation is solved for the run of radial velocity dispersion
where the total force, $g$ is related to the Newtonian force,
$g_N$, by the MOND formula $g\mu(g)=g_N$;  $\mu(x)$ is the 
MOND interpolating function ($\mu(x)=x$ where $x<<1$ and
$\mu(x) = 1$ where $x>>1$), taken here to be of
the form suggested by Zhao \& Famaey (2006).
In each case we adjust the mass of
the sphere in order to reproduce the observed luminosity-weighted 
average line-of-sight velocity within $r_{eff}/2$;
this mass in most cases agrees closely with that implied by the
MOND fundamental plane (eq.\ 5).  The difference between
total (MOND) acceleration and the Newtonian acceleration 
($|g-g_N|$) then
allows us estimate the density distribution of the phantom 
dark halo.  The projected MOND FP mass (X-axis in Fig.\ 2), presumably
the visible mass of the galaxy, is then ``corrected'' by adding
in the projected phantom dark mass.  The result is shown in Fig\ 5
which again shows the lensing mass vs. the MOND FP mass including
the phantom dark matter.  We see that a lensing mass which is
30\% higher than the MOND mass is explained
by the contribution of modified gravity to photon deflection.

We should note, however, that the Zhao-Famaey interpolating function
favours the appearance of phantom dark mass within the optical
image of the galaxy.  This is because the transition from Newton
to MOND is rather more gradual than for the often assumed ('standard')
form of
$\mu$ applied to calculation of galaxy rotation curves 
($\mu(x)=x/\sqrt{1+x^2}$).  Applying the standard form would result in 
a 10\% reduction in the projected phantom dark mass, so the appearance
of Fig.\ 5, and the conclusions we draw from it,
would not be altered.

\section{Conclusions}

With MOND, the baryonic mass-rotation velocity relation for spiral
galaxies,
which forms the basis of the Tully-Fisher law, is exact in so far 
as it relates
to the asymptotic rotation velocity measured far from the luminous
galaxy.  On the other hand, the mass-velocity dispersion relation for
pressure supported systems, the basis of the Faber-Jackson law, 
is only exact for homologous models;  the scaling
of the relation depends upon the detailed characteristics of the object.
Actual elliptical galaxies exhibit a 
range of properties-- various shapes, varying degrees of deviation from
an isothermal state and, no doubt, isotropy of the velocity dispersion--
and cannot be represented by a single homologous sequence of models.
Therefore, spheroidal
galaxies will inevitably present a Faber-Jackson law with 
considerable scatter.  None-the-less, MOND provides an explanation
for the remarkable fact that self-gravitating, pressure-supported 
quasi-isothermal objects with a velocity dispersion of a few
hundred kms$^{-1}$ will have a mass in the range of galaxies-- or objects
with a velocity dispersion $<10$ kms$^{-1}$ will have the mass of globular
clusters-- or objects with 1000 kms$^{-1}$ will have the mass of a
cluster of galaxies.  

In spite of the scatter in the mass-velocity dispersion relation,
when an additional parameter
is added, such as effective radius or surface brightness, MOND models
for elliptical galaxies define a narrow fundamental plane which is
close 
to that implied by the Newtonian virial relation for homologous
objects (isotropic Jaffe models).  This was
not part of the original set of MOND predictions but became
apparent when it was realized that normal elliptical galaxies
are essentially Newtonian systems within the effective radius
and exhibit a wide dispersion in the effective radius-velocity
dispersion relation.
The properties of this fundamental plane were outlined by
a set of 360 large $n$ polytropic spheres with radially
dependent anisotropy chosen to match the observed
joint distribution of ellipticals by effective radius and velocity 
dispersion (Sanders 2000).  Applying this fundamental plane relation
to determine the mass
of those ellipticals in the sample of J{\o}rgenson et al. (1995)
yielded reasonable
values for the mass-to-light ratios. 

Now, thanks to the work of Bolton et al. (2007) we can compare
this mass-based MOND fundamental plane directly to the observed
mass-based fundamental plane as defined by this set of 36 strong
gravitational lenses.  Figs. 1 and 2 illustrate that the
two coincide apart from a systematic offset of about 30\%.
Indeed, the implied MOND mass-to-light ratios are completely
consistent with population synthesis models (Fig.\ 3), and
the small discrepancy between the lensing mass and the
MOND FP mass can be understood in terms of the contribution
of modified gravity to the deflection of photons (Fig.\ 5).
It is important to recall that the properties of the MOND 
fundamental plane (Sanders 2000) were defined well before
those of observed mass-based fundamental plane (Bolton et al. 2007),
so this does, properly speaking, constitute a prediction that
has been subsequently confirmed.  

Most significantly, there
is no evidence from strong gravitational lensing for a significant
mass discrepancy within these high surface density systems-- as MOND
would robustly predict.  This is in contrast to a recent claim by
Ferreras et al. (2008) based upon lensing by six early type galaxies.
They note that the lensing mass, as determined either by General
Relativity or MOND (as extended by TeVeS), is significantly greater
than the stellar mass estimated via population synthesis models.
However, this conclusion appears to give much weight to 
the precision of such models; the mass difference is generally 
smaller than the differences due to the assumption of 
different initial mass functions ($\approx 0.2-0.3$ dex). Moreover, in 
the near infrared, the scatter induced by metallicity effects can be
comparable (Bell et al. 2003).  Overall it is difficult to 
argue that implied mass-to-light ratios 
ranging from two to eight constitute compelling evidence for 
dark matter in early-type galaxies.  

The conclusions here are generally consistent with those of
Zhao et al. (2006).  Looking at a sample of 18 strong
double-image lenses from
the CASTLES sample in the context of TeVeS, they concluded that the implied
lensing masses are consistent with the observed baryonic matter
(with additional deflection provided by the modified gravity).
There are two extreme outliers, but this is not unsurprising in
view of the possible contribution of surrounding groups or
clusters to deflection of photons (in one case, the implied lensing
mass is significantly less than the estimated stellar mass which
would also be a problem for GR).  Treu \& Koopmans (2004)
have argued for the presence of a significant fraction of
dark matter ($f_{DM} = 0.3-0.7$) within the Einstein ring radius
of five well-observed high redshift lens systems (for the 
systems considered
in this paper interpreted in terms of dark matter 
$f_{DM} = 0 - 0.4$).  We have argued here that such a result is expected
because of the larger contribution of modified gravity to
the deflection of photons in higher redshift systems
where the Einstein ring radius is a larger fraction of the
MOND transition radius (Fig.\ 4). 

The results here for this homogeneous relatively nearby sample of
lenses are encouraging for MOND-- primarily with
respect to the basic prediction of small
discrepancy in high surface density systems and secondarily
in confirming
the derived MOND fundamental plane.  Such observations
clearly constitute a powerful test of MOND as an acceleration
based modification. Examples of isolated HSB galaxies showing
a large discrepancy within the bright regions, would be  problematic
for the theory.
 
We thank L\`eon Koopmans and Adam Bolton for sharing
their data on 
the SLACS gravitational lenses.  We are also very grateful to
L\`eon and to Moti Milgrom for helpful comments on the initial 
manuscript.

\end{document}